\def   \ni {\noindent}

\def   \ssk {\vskip  5truept}

\def   \bsk {\vskip 15truept}
 
\def   \newpage {\vfill\eject}
\def   \newline {\hfil\break}

\documentstyle[epsfig]{article}
\begin{document}

\hsize 5truein
\vsize 8truein
\font\abstract=cmr8
\font\keywords=cmr8
\font\caption=cmr8
\font\references=cmr8
\font\text=cmr10
\font\affiliation=cmssi10
\font\author=cmss10
\font\mc=cmss8
\font\title=cmssbx10 scaled\magstep2
\font\alcit=cmti7 scaled\magstephalf
\font\alcin=cmr6 
\font\ita=cmti8
\font\mma=cmr8
\def\ref{\par\noindent\hangindent 15pt}
\null
%\vskip 3.0truecm
%\baselineskip = 12pt

% ------ beginning of font "title" ------

\title{\ni MEV SYNCHROTRON BL LACS}

% beginning of font "author and affiliation"
\bsk \bsk
\author{\ni Gabriele Ghisellini}
\bsk
\affiliation{ Osservatorio Astronomico di Brera, V. Bianchi 46, I-23807 Merate, 
Italy}
                                                
\bsk
\baselineskip = 12pt

% beginning of font "abstract and keywords"
\abstract{ABSTRACT 
The recent BeppoSAX observations of the BL Lac objects
Mkn 501 and 1ES 2344+514 have shown that the synchrotron
spectrum of these objects peaks, in a $\nu$--$\nu F_{\nu}$
representation, at energies at or above 100 keV.
One can wonder if these are the most extreme examples
of hard synchrotron blazars, or if they are the first
cases of a more numerous class of sources.
Here I propose the existence of a class of even more 
extreme BL Lac objects, whose synchrotron spectrum peaks 
at or above 1 MeV.
Based on the observational trend found between the location 
of the synchrotron peak and the bolometric power of BL Lac 
objects, it is argued that the proposed extreme sources 
could have escaped detection (in any band) so far, or 
could have been classified as galaxies, and their 
``BL Lac-ness" could be revealed by INTEGRAL.}                                                    
\bsk
\baselineskip = 12pt

\bsk
\baselineskip = 12pt

% beginning of font "text"

\text{\ni 1. INTRODUCTION
\ssk
\ni     
The blazar class of sources is formed by BL Lacertae 
objects (BL Lacs) and flat spectrum radio quasars (FSRQ),
either with high (HPQ) or low (LPQ) polarization in the optical band.
The differences among subclasses of blazars reflect
the presence or absence of emission lines,
the degree of optical polarization, the band (i.e. radio or X--rays)
where they were discovered.

As an example, consider BL Lacs discovered through radio and X--ray 
surveys: they show different radio to X--ray spectra, but they share
other properties such as the absence of strong emission lines,
the rapid and large amplitude variability and the same average X--ray 
luminosity.
This led Maraschi et al. (1986) and Ghisellini \& Maraschi (1989) to
propose that the spectral differences were due only to the different
viewing angle under which we observe an accelerating, inhomogeneous jet.

On the other hand, Giommi \& Padovani (1994) noticed that the 
spectral energy distribution (SED)
of radio and X--ray selected BL Lacs showed peaks 
(in a $\nu$--$\nu F_\nu$ representation) at different energies, 
and suggested that this difference was intrinsic, and not due to 
orientation effects.
They therefore introduced the notation of HBL (high energy peak BL Lac)
and LBL (low energy peak BL Lac), the former being sources preferentially
selected through X--ray surveys, and the latter through radio surveys.
A crucial help to understand blazars came from 
EGRET, onboard CGRO, and from the ground based Cherenkov telescopes,
such as Whipple and HEGRA: we now know that most of the power emitted
by blazars often lies in the $\gamma$--ray region of the spectrum.
Their SED presents two peaks, the first at mm to X--ray 
\newpage 
\noindent
energies, 
and the second in the MeV to the TeV band.
The energies of the two peaks correlate, in the sense that if the first peak
is located in the mm band, the second is at MeV energies,
while if the first peak is in the X--ray band, the second is
at GeV--TeV energies.
The idea of Padovani \& Giommi (moving peak) can therefore
be extended also to the second peak.

Understanding what rules the SED of blazars is one of the main goal
of the current blazar research.
This is not an easy task, however, since there is still discussion
about the nature of the radiation we see: while the emission
from the radio to UV (i.e. the first peak) should be due to the incoherent
synchrotron process, the emission at higher energies (second peak) could be 
pure Synchrotron Self Compton (SSC: Bloom \& Marscher 1996;
Maraschi, Ghisellini \& Celotti 1992) or a mixture of SSC plus a 
contribution by inverse Compton scattering off photons produced externally 
to the jet (EC: Sikora, Begelman \& Rees 1994; Dermer \& Schlickeiser 1993;
Blandford \& Levinson 1995, Ghisellini \& Madau 1996), 
or another more energetic synchrotron component 
(as in the `proton blazar' model by Mannheim 1993).
In this framework, objects presenting extreme properties are the most
interesting ones, since they best constrain our models.
In this respect the discovery made by BeppoSAX of an extraordinary hard 
X--ray state of the BL Lac objects Mkn 501 and 1ES 2344+514 is extremely
important, and stimulates new and interesting ideas on the physics of 
relativistic jets.

In this paper I will review some recent work on the SED of blazars, 
both from the phenomenological and from the theoretical point of view. 
Then I will discuss the possibility that an entire
new class of BL Lacs can exist, characterized by a synchrotron component
(i.e. the first peak of the SED) peaking at MeV energies.
%Even if some of them can be already present in deep X--ray surveys, 
%these objects 
%could have been mis-classified as normal radio--weak elliptical galaxies.

\bsk
\ni 2. THE BLAZAR SEQUENCE
\ssk
\ni 
Fossati et al. (1998), by considering flux limited samples of 
BL Lac objects and flat spectrum quasars, have noted an interesting
relation between the SED of blazar and their bolometric observed luminosity.
This is shown in Fig. 1.
In less powerful objects the synchrotron peak is at higher energies,
reaching the soft-medium X--ray range.
Analogously, the high energy peak, believed to be due to the inverse
Compton process, shifts to lower energies as the total power increases.
In addition, the ratio between the $\gamma$--ray power and the optical--UV
luminosity increases as the total power increases.

\begin{figure}
\centerline{\psfig{file=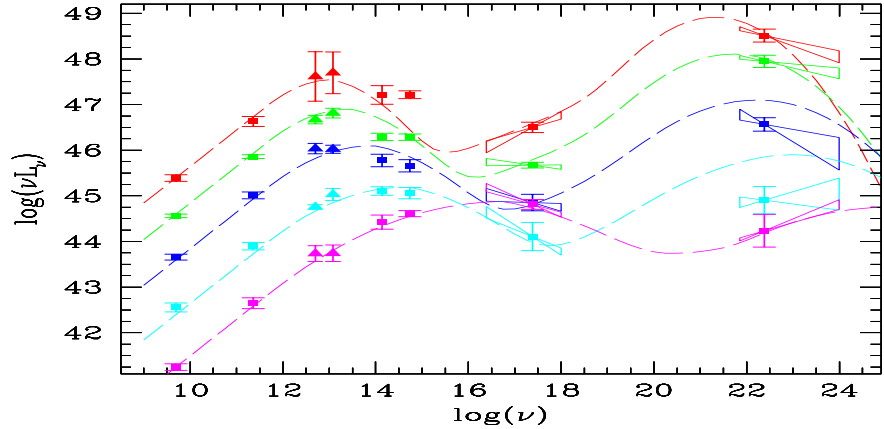, width=13cm, height=7.5truecm}}
\vskip -1 true cm
\caption{FIGURE 1. The average spectra of blazars,
according to their bolometric luminosity, assumed to be traced by the 
radio luminosity. 
Blazars belonging to complete samples have been
devided in 5 luminosity bins, irrespective of their classification.
Notice how the average SED changes as the overall power changes.
Dashed lines corresponds to an analytical, phenomenological fit to the data.
From Fossati et al. 1998.
}
\end{figure}
\begin{figure}
\centerline{\psfig{file=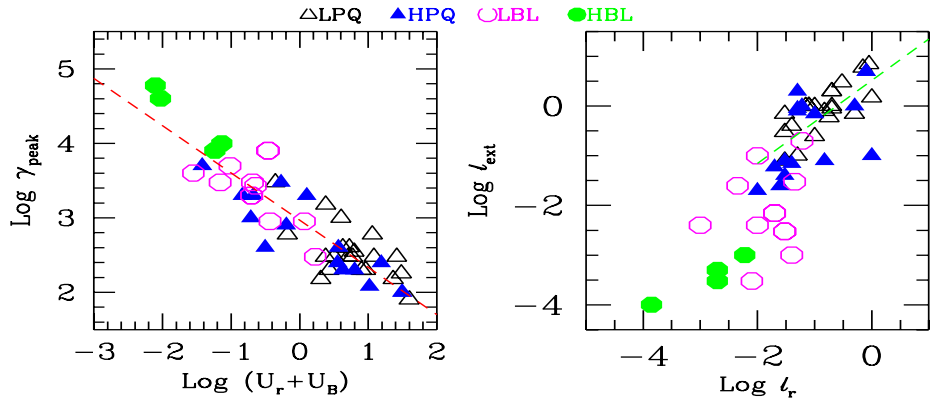, width=13cm, height=10truecm}}
\vskip -4 true cm
\caption{FIGURE 2. Correlations found when modelling the SED of
EGRET blazars with an homogeneous EC model.
The left panel shows the correlation between
$\gamma_{peak}$ and the total (magnetic plus radiative, including
the contribution from external photons) energy density.
Different symbols indicate different classes of blazars, as labelled.
The right panel indicate the correlation between the compactness
($\ell\equiv L\sigma_{\rm T}/(Rm_{\rm e}c^3)$ of the external photons and the 
one corresponding to the injected power, as measured in the comoving
frame of the emitting region. From Ghisellini et al. 1998. }
\end{figure}

Ghisellini et al. (1998), after modelling all blazars detected by EGRET
with an homogeneous EC model, interpreted these results on the basis
of the underlying correlation between the Lorentz factor of the electrons
emitting at the peak, $\gamma_{\rm peak}$, and the amount of energy density
(both magnetic and radiative) present in the emitting region
(see Fig. 2).
This in turn correlates with the observed (beamed) luminosity, and the ratio
between the power of the Compton to the synchrotron components.
Blazars form a sequence: low luminosity lineless BL Lacs (HBL) have
large values of $\gamma_{\rm peak}$, synchrotron peak energy in the 
EUV--soft X--rays and a roughly equally powerful Compton
component peaking in the GeV--TeV band.
LBL are characterized by a greater overall luminosity, a smaller
$\gamma_{\rm peak}$ and peak energies in the optical and GeV band.
More powerful sources, such as HPQ and LPQ, have the smallest values
of $\gamma_{\rm peak}$, peak energies in the mm--far IR and MeV band,
a dominating Compton component in which photons produced externally
to the jet are more important (as seed photons to be Compton
scattered at high energies) than the locally produced synchrotron photons
(see Fig. 2).

%       Text of 2nd paragraph

\bsk
\ni
2.1 The case of Mkn 501 and 1ES 2344+514
\ssk
\ni
In April 1997, BeppoSAX observed Mkn 501 (Pian et al. 1998), one of the 
closest BL Lac objects ($z=0.034$), and the second extragalactic 
source detected in the TeV energy range by Cherenkov telescopes 
(Quinn et al. 1996; Bradbury et al. 1997).
Results of the X--ray observations were extraordinary (Pian et al. 1998,
see also Pian et al., these proceedings), since BeppoSAX detected the source 
in an extremely high and hard synchrotron state, simultaneously with a 
TeV flare. 
The $\nu F_\nu$ synchrotron spectrum of this source was observed to peak
at 100 keV or beyond, while, at $\sim$0.5 TeV, Mkn 501 
was a factor 4--6 brighter than the Crab 
(Catanese et al. 1997, Aharonian et al. 1997).
A similar hard X--ray spectrum was observed by BeppoSAX also for
another TeV source, 1ES 2344+514 (Giommi, Padovani \& Perlman, 1998).
Extreme objects like Mkn 501 and 1ES 2344+514 could be sources where
the particle acceleration mechanism operates at its maximum efficiency, 
succeding to accelerate electrons up to 10 TeV or more.

Mkn 501 and 1ES 2344+514 are probably in the extremely hard state
observed by BeppoSAX only when flaring.
However, there could be other sources that are extremely hard
even in quiescence.
The above discussed trends suggest that these sources should be 
at the lower luminosity end of the BL Lac luminosity function.
It is interesting then to discuss what can limit the relevant electron 
energy $\gamma_{\rm peak}$, which in turn determines where the 
synchrotron spectrum peaks.

\bsk
\ni
3. LIMITS TO THE ELECTRON ENERGIES
\vskip 0.3 true cm
\ni
3.1 Shocks
\ssk
\ni
Guilbert, Fabian \& Rees (1983) derived a useful limit on the maximum
synchrotron frequency that can be produced by shock--accelerated electrons.
In relativistic shocks the Lorentz factor fractional change of the 
electrons for every passage through the shock can be of order unity
($\Delta\gamma/\gamma \sim 1$), with the acceleration timescale
approximately equal to the gyroperiod ($\propto \gamma B^{-1}$, where
$B$ is the magnetic field).
The maximum energy $\gamma_{\rm max}$ is attained when the synchrotron cooling 
timescale $\propto \gamma^{-2}B^{-2}$ equals the acceleration timescale.
Then $\gamma_{\rm max}\propto B^{-1/2}$, resulting in a $B$--independent
maximum synchrotron frequency of 70 MeV.
Additional Compton losses would of course lower this value, but not 
severely, at least in a pure SSC model, since they are inhibited by the 
decline with energy of the Klein--Nishina scattering cross section.

With a Doppler factor $\delta\sim 10$ the observed maximum frequency 
could be $\sim 700$ MeV.
Values of this order lie within the EGRET capability.
Therefore it is unlikely that BL Lacs have synchrotron spectra reaching
energies greater than $\sim$100 MeV.
There must then be a more severe limit to the maximum observable
synchrotron frequency.

\bsk
\ni
3.2 Global energetics
\ssk
\ni
Another limit to the acceleration of particles may come from the
total available power.
This is likely in the form of bulk kinetic motion of the emitting
plasma, $L_{\rm k}$, and in Poynting flux, $L_{\rm B}$. They are given by
(see e.g. Celotti \& Fabian 1993, Ghisellini \& Celotti 1998):

\begin{equation}
L_{\rm k}\, =\, \pi R^2 \Gamma^2 \beta c \, n^\prime\, m_{\rm e} c^2
(<\gamma>+ m_{\rm p}/m_{\rm e});\quad
%\end{equation}
%\begin{equation}
L_{\rm B}\, =\, {1 \over 8} R^2 \Gamma^2 \beta c B^2, 
\end{equation}
where $R$ is the cross sectional radius of the jet,
$n^\prime=\Gamma n$ is the comoving particle density of average energy
$<\gamma>m_{\rm e}c^2$, and $m_{\rm p}$, $m_{\rm e}$ 
are the proton and electron rest masses, respectively. 
An electron proton plasma is assumed.
The synchrotron intrinsic power is 
\begin{equation}
L_{\rm s}^\prime = Volume\,
\int n^\prime (\gamma)\dot \gamma_{\rm s}m_{\rm e}c^2\, d\gamma\, =\,
{2\over 9}\, R^3 \sigma_{\rm T} c n^{\prime} B^2 <\gamma^2>
\end{equation}
where
$\dot\gamma_{\rm s}$ is the synchrotron cooling rate, and 
$n^\prime (\gamma)\propto \gamma^{-p}$ between
$\gamma_{\rm min}$ and $\gamma_{\rm peak}$.
For a viewing angle $\sim 1/\Gamma$, the luminosity calculated assuming 
isotropy is related to $L_{\rm s}^\prime$ by 
$L_{\rm s,obs}=\Gamma^4 L_{\rm s}^\prime$.
The intrinsic power emitted over the entire solid angle equals
$\Gamma^2L_{\rm s}^\prime$.
We can then relate the synchrotron power $\Gamma^2 L_{\rm s}^\prime$  
to $L_{\rm k}$ (which is proportional to $n^\prime$)
and $L_{\rm B}$ (which is proportional to the magnetic energy density 
$U_{\rm B}$),
obtaining
\begin{equation}
\Gamma^2 L_{\rm s}^\prime \, =\, 
{16 \over 9\pi} \, \, {\sigma_{\rm T} L_{\rm k} L_{\rm B} 
\over R m_{\rm e}c^3 \Gamma^2}
\, \, { <\gamma^2> \over <\gamma>+m_{\rm p}/m_{\rm e}}
\end{equation}
Requiring $ \Gamma^2 L_{\rm s}^\prime < L_{\rm k}$ implies:
\begin{equation}
{ <\gamma^2> \over <\gamma>+m_{\rm p}/m_{\rm e} } \, <\, 
{9\pi \over 16}\, {R m_{\rm e}c^3 \Gamma^2 \over \sigma_{\rm T}  L_{\rm B}}
\end{equation}
At high energies, the synchrotron process is efficient and fast,
and a quasi--steady emission requires continuos acceleration of
particles, at a rate that balances the radiative (synchrotron) losses.
As discussed by Ghisellini \& Celotti (1998), the most efficient
synchrotron emitting jet corresponds to equipartition
between bulk kinetic and Poynting {\it powers}. 
The observed synchrotron power is then maximized for 
$L_{\rm k}\sim L_{\rm B}$. 
Setting $\Gamma^2 L_{\rm s}^\prime$ at its 
maximum possible value 
($\Gamma^2 L_{\rm s}^\prime\sim L_{\rm k}\sim L_{\rm B}$),
assuming $n(\gamma)\propto \gamma^{-2}$ 
between $\gamma_{\rm min} < m_{\rm p}/m_{\rm e}$ and $\gamma_{\rm peak}$
we have a limit for $\gamma_{\rm peak}$:
\begin{equation}
\gamma_{\rm peak}\, <\, 
{9\pi \over 16\gamma_{\rm min}}\, {R m_{\rm p}c^3  \over \sigma_{\rm T}}\,
{1 \over L_{\rm s}^\prime}\, =\, 
1.2\times 10^6\, {R_{16}\over \gamma_{\rm min}L^\prime_{\rm s,42}}
\end{equation}
Here the notation $Q=10^xQ_x$ is used, with cgs units.
The observed synchrotron peak frequency
$\nu_{\rm peak}=3.7\times 10^6 \gamma_{\rm peak}^2B\Gamma$  Hz corresponds to
\begin{equation}
\nu_{\rm peak} \, =\, 0.36\, {R_{16} \Gamma_1 \over 
\gamma_{\rm min}^2(L^\prime_{s,42})^{3/2}} \quad {\rm MeV}
\end{equation}
If only a fraction $a$ of the bulk motion power is transformed
into $\Gamma^2 L_{\rm s}^\prime$, then the above estimate must 
be multiplied by $a^{7/2}$.
From Eq.(6)  we have that the 1997 flare of
Mkn 501, with $L^\prime_{\rm s}\sim 2\times 10^{42}/\Gamma_1^4$ erg s$^{-1}$
and $\nu_{\rm peak}\sim$100 keV, 
was quite close to transform all the available bulk motion
power into synchrotron emission.
For $\Gamma=15$ and $R=10^{16}$ cm, its emitted synchrotron power 
is $\sim$40\% of $L_{\rm k}$.

The above arguments suggest that $\sim$1 MeV can be considered
a limit for the observed maximum frequency of BL Lacs.
If $a$ is relatively constant, we also have an inverse correlation 
between $\nu_{\rm peak}$ and $L^\prime_{\rm s}$, in the sense that
the most extreme objects in terms of $\nu_{\rm peak}$ are
the least powerful ones.

\bsk
\ni
4. PREDICTED SPECTRA
\ssk
\ni
Figs 3 and 4 illustrate the possible spectrum of a MeV BL Lac, assuming
as a working hypothesis an homogeneous one--zone SSC model.
Relativistic electrons are assumed to be continuously injected throughout
a spherical source of size $R$ embedded in a tangled magnetic field $B$.
The injected particles have a power law energy distribution
$\propto \gamma^{-s}$ between $\gamma_{\rm min}$ and $\gamma_{\rm max}$.
The steady emitting particle distribution is found by solving a 
continuity equation, i.e. by balancing injection and radiative cooling.
The Klein--Nishina decline of the scattering cross section and the
possible photon-photon e$^\pm$ pair production are accounted for.
Details on the model can be found in Ghisellini (1989) and
Ghisellini et al. (1998).

\begin{figure}
\centerline{\psfig{file=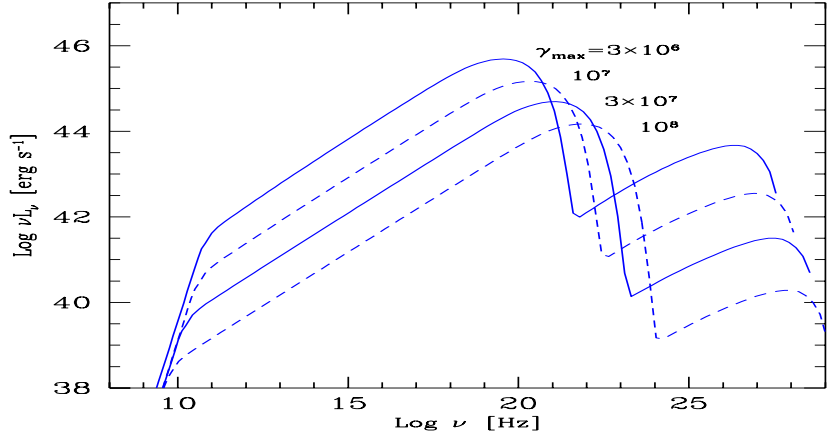, width=13cm, height=7truecm}}
\vskip -0.5 true cm
\caption{FIGURE 3. Some synchrotron self Compton spectra obtained by 
continuously injecting electrons up to different $\gamma_{\rm max}$,
as labelled, corresponding to different total luminosities.
From Ghisellini et al., in preparation.}
\end{figure}

\begin{figure}
\centerline{\psfig{file=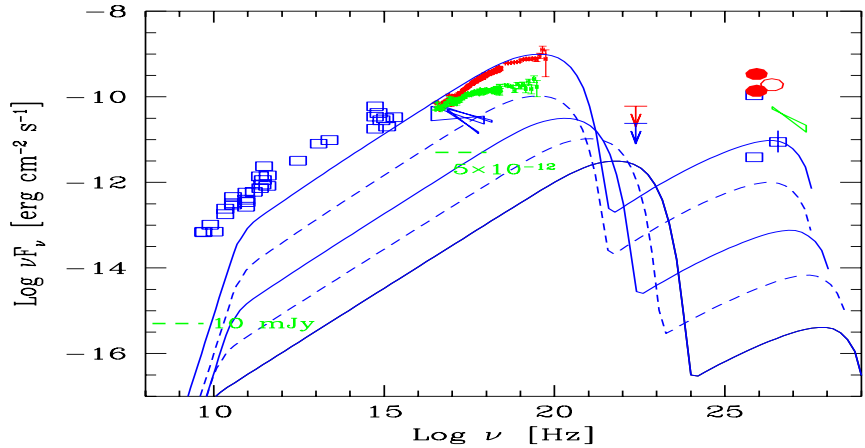, width=13cm, height=7truecm}}
\caption{FIGURE 4. Same spectra as of Fig. 3, 
here shown in the $\nu$--$\nu F_{\nu}$ representation, assuming a redshift
$z=0.1$ for all models but the top one, which is the same model
as the second one, but for $z=0.034$ (redshift of Mkn 501).
The dashed horizontal lines marks the level of 10 mJy at 5 GHz,
approximately the level of the faintest X--ray selected BL Lacs,
and of $5\times 10^{-12}$ erg s$^{-1}$ between 0.3 and 3.5 keV,
approximately the limit flux of the Einstein SLEW survey.  
The SED of Mkn 501 is shown for comparison.
From Ghisellini et al., in preparation.}
\end{figure}

Also shown in Fig. 4 is the spectrum of Mkn 501 of the
flaring state of April 1997, together with other, non simultaneous, data.
The computed SSC models obey the
$\gamma_{\rm peak}\propto L_{\rm s}^{-1}$ relation of Equation (6).
The first of these models (top curve) assumes
$\gamma_{\rm max}=3\times 10^6$, $L_{\rm s,obs} = 2\times 10^{46}$ erg s$^{-1}$
$R=10^{16}$ cm, $\delta=15$ and $B=0.9$ Gauss.
The intrinsic injected power is 
$L^\prime_{\rm inj}= 5.5\times 10^{41}$ erg s$^{-1}$.
The second model (second line from the top) is the same, but
the assumed redshift is $z=0.1$, and this redshift is assumed also
for the remaining models.
The size and the beaming factor are the same for all models.
The magnetic field scales as $L^{1/2}$, and each intrinsic 
(as well as observed) luminosity differs for a factor 3, with
$\gamma_{\rm max}$ varying accordingly.
All models assume
$I_{obs}(\nu_{obs}) =\delta^3 I^\prime(\nu^\prime)$
for the transformation between the observed and the intrinsic intensity.

It can be seen that the plotted SED are at or below the current
limits of large area X--ray survey, below the approximate sensitivity
limit of EGRET, and the expected radio flux is at level of 1 mJy or less.
But in the MeV region these sources are bright, even if currently
below the COMPTEL capabilities.
Note also that the Compton flux is severely inhibited by Klein--Nishina 
effects. 
However, the Compton flux is calculated assuming scattering between 
electrons and photons produced by the same electron populations;
any additional mm--far infrared photons field can increase the 
emitted high energy flux. 

Note, from Fig. 4, that the 0.1--10 GeV flux is below the
current EGRET sensitivity level, and that, in the optical band,
the contribution of an underlying galaxy could hide the
non--thermal continuum.
In the X--ray band the flux could be strong enough to let these
sources be present in moderately deep X--ray surveys, such as the Rosat RASS
and the Einstein EMSS.
However, note that the radio flux of these sources is at the mJy level,
at or below the limit of the present large area radio sky surveys.
This may be the reasons why we have not yet discovered them, i.e.
some of these sources are too faint at all frequencies but the MeV band,
and even if bright enough in the X--ray band to be included in
current samples, they could have been classified as normal 
(radio--weak) elliptical galaxies.
INTEGRAL could instead discover the brightest objects through their intense
MeV emission.

}

%\bsk
%\ni 5. CONCLUSIONS 
%\ssk
%\ni 

\vskip 0.3 true cm
\baselineskip = 12pt
{\abstract \ni ACKNOWLEDGMENTS

I thank Annalisa Celotti and Laura Maraschi for useful discussions.}
\vskip 0.3 true cm

\baselineskip = 12pt

{\references \ni REFERENCES
\ssk

\ref Aharonian F. et al., 1997, A\&A 327 L5
\ref Blandford, R.D. \& Levinson, A. 1995, ApJ, 441, 79
%\ref Bloom, S.D. \& Marscher, A.P. 1993, in: Proceedings of the Compton
%     Symposium, eds. M. Friedlander \& N. Gehrels (New York: AIP), 578
\ref Bloom, S.D. \& Marscher, A.P. 1996, ApJ, 461, 657
\ref Bradbury, S. M., et al. 1997, A\&A, 320, L5
\ref Catanese M. et al., 1997, ApJ 487, L143
\ref Celotti A. \& Fabian A.C., 1993, MNRAS, 264, 228
\ref Dermer C.D., Schlickeiser R., 1993, ApJ, 416, 458
\ref Fossati, G.,  Celotti A., Maraschi L. Comastri A. \& Ghisellini G., 
    1998, MNRAS, 229, 433
\ref Ghisellini G., 1989, MNRAS, 267, 167
\ref Ghisellini G. \& Maraschi L., 1989, ApJ, 340, 181
\ref Ghisellini G. \& Madau P., 1996, MNRAS, 280, 67
\ref Ghisellini G., Celotti A., Fossati G., Maraschi L. \&
     Comastri A., 1998, MNRAS, in press.
\ref Ghisellini G. \& Celotti A., 1998, submitted to MNRAS
\ref Giommi P. \& Padovani P., 1994, ApJ, 444, 567
\ref Giommi P., Padovani P. \& Perlman E., 1998, in The Active X--ray Sky:
     results from BeppoSAX and Rossi--XTE, Eds. L. Scarsi, Bradt, 
     P. Giommi \& F. Fiore, in press.
\ref Guilbert P.W., Fabian A.C. \& Rees M.J., 1983, MNRAS, 205, 593
\ref Maraschi L., Ghisellini G., Treves A. \& Tanzi E.G., 1986, ApJ, 310, 325
\ref Maraschi L., Ghisellini G., Celotti A., 1992, ApJ, 397, L5
\ref Mannheim K., 1993, A\&A, 269, 67
\ref Padovani P. \& Giommi P., 1995, ApJ, 444, 567 % HBL LBL
\ref Pian E. et al., 1998 ApJ, 
%\ref Protheroe R.J. et al., 25th Cosmic Ray Conf. Durban 1997,
%     astro-ph/9710118
\ref Quinn, J., et al. 1996, ApJ, 456, L83
\ref Sikora M., Begelman M.C., Rees M.J., 1994, ApJ, 421, 153

}

\end{document}